\documentclass[traditabstract,twocolumn]{aa}

\usepackage{epsfig,graphicx,natbib}
\usepackage{longtable}
\usepackage{amssymb}
\def\Tcmb{\hbox{$T_\mathrm{CMB}$}}
\def\Trot{\hbox{$T_\mathrm{rot}$}}

\def\nH2{\hbox{$n_\mathrm{H_2}$}}
\def\kms{\hbox{km\,s$^{-1}$}}
\def\PKS1830{\hbox{PKS\,1830$-$211}}
\def\cm-2{\hbox{cm$^{-2}$}}
\def\fH2{\hbox{$f_{\rm H2}$}}

\begin{document}

\title{Detection of CH$^+$, SH$^+$, and their $^{13}$C- and $^{34}$S- isotopologues toward \PKS1830}

\author{S.~Muller \inst{1}
\and H.\,S.\,P.~M{\"u}ller \inst{2}
\and J.\,H.~Black \inst{1}
\and M.~G\'erin \inst{3}
\and F.~Combes \inst{4}
\and S. Curran \inst{5}
\and E.~Falgarone \inst{3}
\and M.~Gu\'elin \inst{6,3}
\and C.~Henkel \inst{7,8}
\and S. Mart\'in \inst{9,10}
\and K.\,M.~Menten \inst{7}
\and E.~Roueff \inst{3}
\and S.~Aalto \inst{1}
\and A.~Beelen \inst{11}
\and T. Wiklind \inst{12}
\and M.\,A.~Zwaan \inst{13}
}

\institute{Department of Space, Earth and Environment, Chalmers University of Technology, Onsala Space Observatory, SE-43992 Onsala, Sweden
\and I.~Physikalisches Institut, Universit{\"a}t zu K{\"o}ln, Z{\"u}lpicher Str. 77, 50937 K{\"o}ln, Germany
\and LERMA/LRA, Ecole Normale Sup\'erieure, Observatoire de Paris, CNRS UMR 8112, PSL Research University, Sorbonne Universit\'es, UPMC Universit\'e Paris, 24 rue Lhomond, 75005, Paris, France
\and Observatoire de Paris, LERMA, College de France, CNRS, PSL Univ., UPMC, Sorbonne Univ., F-75014, Paris, France
\and School of Chemical and Physical Sciences, Victoria University of Wellington, PO Box 600, Wellington 6140, New Zealand 
\and Institut de Radioastronomie Millim\'etrique, 300, rue de la piscine, 38406 St Martin d'H\`eres, France 
\and Max-Planck-Institut f\"ur Radioastonomie, Auf dem H\"ugel 69, D-53121 Bonn, Germany
\and Astron. Dept., King Abdulaziz University, P.O. Box 80203, Jeddah, Saudi Arabia
\and European Southern Observatory, Alonso de C\'ordova 3107, Vitacura, Santiago, Chile
\and Joint ALMA Observatory, Alonso de C\'ordova 3107, Vitacura, Santiago, Chile
\and Institut d'Astrophysique Spatiale, B\^at. 121, Universit\'e Paris-Sud, 91405 Orsay Cedex, France
\and Department of Physics, Catholic University of America, 620 Michigan Ave NE, Washington DC 20064
\and European Southern Observatory, Karl-Schwarzschild-Str. 2, 85748 Garching b. M\"unchen, Germany
}

\date {Received  / Accepted}

\titlerunning{CH$^+$ and SH$^+$ toward \PKS1830}
\authorrunning{}

\abstract{The $z=0.89$ molecular absorber toward \PKS1830\ provides us with the opportunity to probe the chemical and physical properties of the interstellar medium in the disk of a galaxy at a look-back time of half the present age of the Universe. Recent ALMA observations of hydrides have unveiled the multi-phase composition of this source's interstellar medium along two absorbing sightlines. Here, we report ALMA observations of CH$^+$ and SH$^+$, and of their $^{13}$C- and $^{34}$S- isotopologues, as potential tracers of energetic processes in the interstellar medium. CH$^+$ and $^{13}$CH$^+$ are detected toward both images of \PKS1830, CH$^+$ showing the deepest and broadest absorption among all species observed so far. The [CH$^+$]/[$^{13}$CH$^+$] abundance ratio is $\sim 100$ in the south-west line of sight. This value is larger than any previous [$^{12}$CX]/[$^{13}$CX] ratios determined from other species toward this source and suggests either that the latter might be affected by fractionation or that CH$^+$ might be tracing a different gas component. Toward the north-east image, we find an even larger value of [CH$^+$]/[$^{13}$CH$^+$], $146 \pm 43$, although with a large uncertainty. This sightline intercepts the absorber at a larger galactocentric radius than the southwestern one, where material might be less processed in stellar nucleosynthesis. In contrast to CH$^+$ and its $^{13}$C isotopologue, SH$^+$ and $^{34}$SH$^+$ are only detected on the south-west sightline. These are the first detections of extragalactic SH$^+$ and interstellar $^{34}$SH$^+$. The spectroscopic parameters of SH$^+$ are reevaluated and improved rest frequencies of $^{34}$SH$^+$  are obtained. The [CH$^+$]/[SH$^+$] column density ratios show a large difference between the two lines of sight: $\sim 25$ and $>600$ toward the SW and NE image, respectively. We are not able to shed light on the formation process of CH$^+$ and SH$^+$ with these data, but the differences in the two sightlines toward \PKS1830\ suggest that their absorptions arise from gas with molecular fraction $\gtrsim $10\%, with SH$^+$ tracing significantly higher molecular fractions than CH$^+$.
}
\maketitle

\section{Introduction}

Hydrides are formed by the first chemical reactions starting from atomic gas and are therefore at the root of interstellar chemistry. Due to the relative simplicity of their chemical network, they provide excellent diagnostics of the physico-chemical properties of the interstellar gas, especially of the low-density diffuse component (\citealt{ger16}).

Methylidynium, CH$^+$, was one of the first interstellar molecules to be identified, in absorption in the optical spectrum toward bright stars \citep{dou41}. This discovery established a long lasting problem for astrochemistry, the formation and survival of this reactive species in the interstellar medium (ISM). Its large observed abundances are at odds with predictions from steady-state quiescent gas-phase chemical models by several orders of magnitude. The formation path of the molecule, C$^+$ + H$_2$ $\rightarrow$ CH$^+$ + H, has a high endothermicity of $\Delta E/k \sim 4300$~K, and mechanisms such as shocks (e.g., \citealt{eli80}), dissipation of turbulence (e.g., \citealt{god09, god14}), intense FUV or X-ray radiation (e.g., \citealt{mor16}), or formation from vibrationally excited H$_2$ (e.g., \citealt{agu10, nag13, fau17}) are advocated to overcome this problem. 

The $J$=1-0 transition of CH$^+$ (and its $^{13}$C isotopologue), at 835~GHz (830~GHz, respectively), has been observed in absorption in Galactic sightlines toward massive star-forming regions with the Herschel Space Observatory (\citealt{fal10,god12}). In addition, optical absorption lines of CH$^+$ have been observed toward stars in the Magellanic Clouds (\citealt{wel06}) and even bright supernovae outside the Local Group (SN1986G in NGC5128, \citealt{dodo89};  SN2006X in M100, \citealt{cox08}; SN2008fp in ESO428$-$G14, \citealt{cox14}; SN2014J in M82, \citealt{rit15}), allowing us to probe extragalactic ISM lines of sight.

The sulfanylium ion SH$^+$ has an even higher endothermicity of $\Delta E/k \sim 9900$~K in its formation via the reaction of S$^+$ with ground-state H$_2$. SH$^+$ can also be formed by reactions of S$^+$ with vibrationally excited H$_2(v>1)$ (\citealt{zan13}) and in X-ray dominated regions by reactions of S$^{++}$ with H$_2$ (\citealt{abe08}). SH$^+$ was first detected in space in emission toward the massive star forming region W3\,IRS5 (\citealt{ben10}) with the Herschel Space Observatory and in absorption toward the star-forming region complex Sgr\,B2 near the Galactic Center (\citealt{men11}) with the Atacama Pathfinder EXperiment telescope. Unlike CH$^+$, it has never been detected in an extragalactic source before. 

Because the formation of CH$^+$ and SH$^+$ have different endothermicity, their abundance ratio could reflect the physical properties of the region where these molecules form. \cite{god12} did a comparative study of the CH$^+$ and SH$^+$ absorptions along multiple Galactic sightlines, and found that the CH$^+$/SH$^+$ column density ratios can vary by more than two orders of magnitude, from one to more than 100. They find, however, a relatively good correlation between N(CH$^+$)/N(SH$^+$) and N(CH$^+$)/N(H), except toward the Central Molecular Zone in the Milky Way. There, they argue that the stronger X-ray radiation field could trigger reactions of C$^{++}$ and S$^{++}$ ions with H$_2$ to enhance the abundances of CH$^+$ and SH$^+$.

Here, we report the detection of CH$^+$, SH$^+$, and their $^{13}$C- and $^{34}$S- isotopologues in the $z$=0.89 absorber located in front of the quasar \PKS1830. The quasar, at $z$=2.5 (\citealt{lid99}), is gravitationally lensed by the foreground absorber, a nearly face-on spiral galaxy (\citealt{win02}). The directions to the two bright and compact lensed images of the quasar (separated by 1$\arcsec$) form two independent lines of sight through the disk of the intervening galaxy, with absorption detected for many molecular species (e.g., \citealt{wik96, mul11,mul14a}).

\section{Observations} \label{sec:data}

The observations were carried out with the Atacama Large Millimeter/submillimeter Array (ALMA) during its early cycles (1 and 2).

{\em CH$^+$ tuning:} The CH$^+$ and $^{13}$CH$^+$ $J$=1-0 transitions, redshifted to $\sim$440~GHz in ALMA band~8, were observed simultaneously on 2015 May 20. The weather conditions were excellent, with a precipitable water vapor content of $\sim$0.3~mm. The total on-source time was about 17 minutes. The array was composed of 35 antennas, resulting in a synthesized beam of $\sim$0.3$\arcsec$, full-width at half maximum (FWHM). Hence, the two lensed images of \PKS1830\ were well resolved. The correlator was configured with 1.875~GHz wide spectral windows and a spectral resolution of 1.1~MHz, corresponding to a velocity resolution, after Hanning smoothing, of $\sim$0.8~\kms.

{\em SH$^+$ tuning:} The SH$^+$ and $^{34}$SH$^+$ $N_J$=$1_2$-$0_1$ transitions, redshifted to $\sim$280~GHz in ALMA band~7, were observed on 2014 May 5 (two executions) and July 18 (1 execution). The H$_2^{18}$O $(1_{10}$-$1_{01})$ line (rest frequency 547.676~GHz, redshifted to 290.4~GHz) was also observed with the same tuning. The weather conditions were good to moderate, with a precipitable water vapor content between 0.5--2.5~mm. The total on-source time comprised about 1 hour. In each execution, the array was composed of 30 antennas, resulting in a final synthesized beam smaller than 0.5$\arcsec$ FWHM. The correlator was also configured with 1.875~GHz wide spectral windows and a spectral resolution of 1.1~MHz, corresponding to a velocity resolution, after Hanning smoothing, of $\sim$1.2~\kms.

For both the CH$^+$ and the SH$^+$ tunings, the data calibration was done within the CASA\footnote{http://casa.nrao.edu/} package, following a standard procedure. The bandpass response of the antennas was calibrated from observations of the bright quasar J\,1924$-$292. The gain solutions were self-calibrated on the continuum of \PKS1830. The final spectra were extracted toward both lensed images of \PKS1830\ using the CASA-python task UVMULTIFIT (\citealt{mar14}) and fitting a model of two point sources to the interferometric visibilities.

\section{Laboratory spectroscopic data} \label{sec:spectro}

The spectroscopic parameters for the CH$^+$ and SH$^+$ lines discussed in this paper are given in Table\,\ref{tab:linelist}. The CH$^+$ and $^{13}$CH$^+$ rest frequencies were taken from the Cologne Database for Molecular Spectroscopy (CDMS)~\footnote{http://www.astro.uni-koeln.de/cdms/} \citep{CDMS01,CDMS05}, based on \cite{hmul10}. Both $J$=1-0 transition frequencies were determined by \cite{ama10}. The rest frequencies for the SH$^+$ $N_J$=$1_2$-$0_1$ transitions are taken from laboratory measurements by \cite{hal15}. The CDMS SH$^+$ entry is based on the present work as detailed in the Appendix~\ref{app:SH+}.

In the Born-Oppenheimer approximation, the SH$^+$ spectroscopic data can be taken directly to derive rest frequencies of $^{34}$SH$^+$ \citep{bro86}, see also \cite{hmul15b} for the determination of NO spectroscopic parameters from data of several isotopic variants and references therein for further examples. The resulting $^{34}$SH$^+$ $N$=1-0 transition frequencies are given in Table\,\ref{tab:spectro34SH+}. Deviations from the Born-Oppenheimer approximation are not known accurately but are small in the case of a substitution of $^{32}$S by $^{34}$S. Trial fits with plausible values suggest that the most important correction, the one to the rotational constant $B$, may shift all $N$=1-0 transition frequencies by around 1~MHz, usually, but not always, to higher frequencies. The correction to the spin-spin coupling parameter $\lambda$ may cause shifts of around one to a few megahertz also, but affects mostly the upper and lower frequency FS components in opposite directions, whereas the shift of the $J$=2-1 line is much smaller, possibly several 100~kHz. Other corrections as well as uncertainties from the values and uncertainties of the higher spectroscopic parameters are most likely smaller, but may add up to several 100~kHz.

We adopt the electric dipole moment $\mu_{{\rm CH}^+} = 1.68$ Debye (\citealt{che07}) and note that a nearly identical value $\mu_{^{13}{\rm CH}^+} = 1.7$ Debye was reported for the heavier isotopologue (\citealt{fol87}). For the sulfanylium ion we adopt the same value $\mu_{{\rm SH}^+}=1.28$ Debye for both isotopologues (\citealt{sen85,che07}).

\begin{table*}[ht]
\caption{Line parameters.}
\label{tab:linelist}
\begin{center} \begin{tabular}{lcccccc}
\hline \hline
\multicolumn{1}{c}{Line} & Observation & Rest freq. & Sky freq. & $E_{low}$ $^{(a)}$ & $S_{ul}$ & $\alpha$ $^{(c)}$   \\
                         & date        &(GHz)      & (GHz)     & (K)              & $^{(b)}$ & ($10^{12}$~\cm-2\,km$^{-1}$\,s) \\

\hline

SH$^+$ $N_J$=$1_2$-$0_1$ $F$=1.5-0.5 & 2014 May 05, Jul 18 & 526.038793(50) & 278.944 & 0.0 & 1.14 &          \\ 
    \multicolumn{1}{r}{$F$=2.5-1.5} & -- & 526.048023(50) & 278.949 & 0.0 & 2.06 &   30.1 \\
    \multicolumn{1}{r}{$F$=1.5-1.5} & -- & 526.124951(50) & 278.990 & 0.0 & 0.23 &      \\

$^{34}$SH$^+$ $N_J$=$1_2$-$0_1$  & -- & \multicolumn{5}{c}{see dedicated entries in Table\,\ref{tab:spectro34SH+}} \\

\hline

CH$^+$ $J$=1-0       & 2015 May 20 & 835.137504(20) & 442.851 & 0.0 & 1.0 & 2.85 \\
$^{13}$CH$^+$ $J$=1-0 & --  & 830.216096(22) & 440.241 & 0.0 & 1.0 & 2.79  \\

\hline

\end{tabular} 
\tablefoot{Frequencies and molecular data are taken from the Cologne Database for Molecular Spectroscopy (\citealt{CDMS01}), except for $^{32}$SH$^+$, for which they are taken from direct laboratory measurements by \cite{hal15}. See also Sec.\ref{sec:spectro}. The sky frequencies are calculated using z=0.88582 (heliocentric frame). \\
$(a)$ $E_{low}$ is the lower-level energy of the transition. $(b)$ $S_{ul}$ is the line strength. 
$(c)$ The coefficient $\alpha$ is defined in Eq.\,\ref{eq:alpha} and calculated under the assumption that the rotational excitation is coupled with the cosmic microwave background temperature, 5.14~K at $z$=0.89 (see, e.g., \citealt{mul13,mul14a}). In case of hyperfine structure (SH$^+$), the coefficient applies to the hyperfine-structure deconvolved line profile, normalized for an equivalent component with $S_{ul}=1$.
}
\end{center} \end{table*}

\begin{table*}
\caption{Quantum numbers, frequencies, Einstein $A$ coefficients, upper $g_u$ and lower $g_l$ state degeneracies, and upper $E_u$ and lower $E_l$ state energies of the $N$=1-0 ground state rotational transition of $^{34}$SH$^+$.}
\label{tab:spectro34SH+}
\begin{center}
\begin{tabular}{ccccccccc}
\hline \hline
$N' - N''$ & $J' - J''$ & $F' - F''$ & Frequency $^a$ & $A$                 & $g_u$ & $g_l$ & $E_u$ & $E_l$ \\
           &            &            & (MHz)          & (10$^{-4}$~s$^{-1}$) &            &            & (cm$^{-1}$)  & (cm$^{-1}$) \\ 
\hline
1--0 & 0--1 & 0.5--0.5 & 344896.7 &  1.14 & 2 & 2 & 11.5074 &  0.0029 \\
     & 0--1 & 0.5--1.5 & 344982.8 &  2.28 & 2 & 4 & 11.5074 &  0.0000 \\
     & 2--1 & 1.5--0.5 & 525085.2 &  7.95 & 4 & 2 & 17.5178 &  0.0029 \\
     & 2--1 & 2.5--1.5 & 525094.4 &  9.54 & 6 & 4 & 17.5153 &  0.0000 \\
     & 2--1 & 1.5--1.5 & 525171.4 &  1.59 & 4 & 4 & 17.5178 &  0.0000 \\
     & 1--1 & 1.5--0.5 & 682382.5 &  2.88 & 4 & 2 & 22.7647 &  0.0029 \\
     & 1--1 & 0.5--0.5 & 682408.4 & 11.53 & 2 & 2 & 22.7655 &  0.0029 \\
     & 1--1 & 1.5--1.5 & 682468.6 & 14.42 & 4 & 4 & 22.7647 &  0.0000 \\
     & 1--1 & 0.5--1.5 & 682494.5 &  5.77 & 2 & 4 & 22.7655 &  0.0000 \\

\hline
\end{tabular}
\end{center}
\tablefoot{$^a$ Uncertainties of at least $\sim$2~MHz mostly because of a breakdown of the Born-Oppenheimer approximation, see Sec.\,\ref{sec:spectro}.}
\end{table*}

\section{Results}

In this section, we present the absorption spectra of CH$^+$, $^{13}$CH$^+$, SH$^+$, and $^{34}$SH$^+$, obtained with ALMA toward the two lensed images of \PKS1830. The spectra were first normalized to the continuum level of each image and are described as:

\begin{equation}
I_{abs}(v) = 1- f_c \times \left [ 1-\exp{\{- \sum _i \tau_i (v)\}} \right ],
\label{eq:abs_function}
\end{equation}
\noindent where $f_c$ is the continuum source covering factor (i.e., the fraction of the continuum emission actually covered by the absorbing clouds) which can vary for different species, and $\tau_i(v)$, the optical depths of different velocity components, which we assume to have Gaussian profiles. In view of the following sections and Figs.\,\ref{fig:specSW}--\ref{fig:fitNE}, this choice of Gaussian profiles is a good approximation.

The size of the continuum emission at submillimeter wavelengths is as yet unknown, but Very Long Baseline Interferometry (VLBI) measurements at 7~mm indicate a size of $\sim$0.2~mas for the southwest image, scaling linearly with $\lambda$ (\citealt{jin03}). Projected in the plane of the absorber, the apparent sizes of the submm continuum images are thus $\sim$0.1~pc in diameter. Since ALMA cannot spatially resolve this scale, the degeneracy between $f_c$ and optical depth can only be broken for a heavily saturated line (as for CH$^+$), or possibly if spectrally resolved multiple fine/hyperfine components of different strengths allow us to constrain their opacity. In the absence of such constraints, we assume $f_c$=1 to derive apparent optical depths, and thus obtain lower limits to the column densities. From previous ALMA observations, \cite{mul14a} observed $f_c$(SW) between 91\%-95\%, with a trend of increasing $f_c$ with increasing frequency (i.e., when the size of the continuum emission becomes smaller). However, chemical segregation and time variations of the continuum morphology complicate the picture. The covering factor is unknown for the NE image, the OH$^+$ absorption observed by \cite{mul16} implying $f_c$(NE)$>$50\% in June 2015.

Next, we derive column densities by assuming that the excitation of CH$^+$ and SH$^+$ is strongly coupled to the cosmic microwave background (CMB, \Tcmb=5.14~K at $z$=0.89, \citealt{mul13}) and that other contributions to the excitation are small. Although rotationally inelastic collisions with H, H$_2$, and $e^-$ might need to be considered, reactive collisions of CH$^+$ with the same collision partners are even faster, so that the formation and destruction processes must be considered in proper analysis of the excitation (cf. \citealt{god13}). Departures from the CMB can also be caused by radiative excitation in the local background radiation of the absorbing galaxy itself. We have no information about the local visible, infrared, and submm-wave continuum inside the absorbing galaxy. A local continuum comparable to that of the average background in the Milky Way would dominate the excess excitation of CH$^+$ in competition with collisions at the densities and temperatures of a Galactic diffuse molecular cloud. Even so, the excitation temperature in CH$^+$ $J$=1-0 might be raised to $\sim 6$~K, which would still have a negligible effect on the following analysis. Local excitation effects are unlikely to be any larger in SH$^+$ under diffuse-cloud conditions. Accordingly, we calculate the column densities as:
\begin{equation} \label{eq:alpha}
N_{col} = \alpha_{ij} \times \int \tau_{ij} dv,
\end{equation}
\noindent where the $\alpha_{ij}$ coefficients (Table\,\ref{tab:linelist}) are calculated for a given transition $i$-$j$ for \Trot=5.14~K (see e.g., \citealt{mul14a}).

\subsection{\PKS1830\ south-west line of sight}

\subsubsection{CH$^+$ and $^{13}$CH$^+$}

\begin{figure}[t!] \begin{center}
\includegraphics[width=8.8cm]{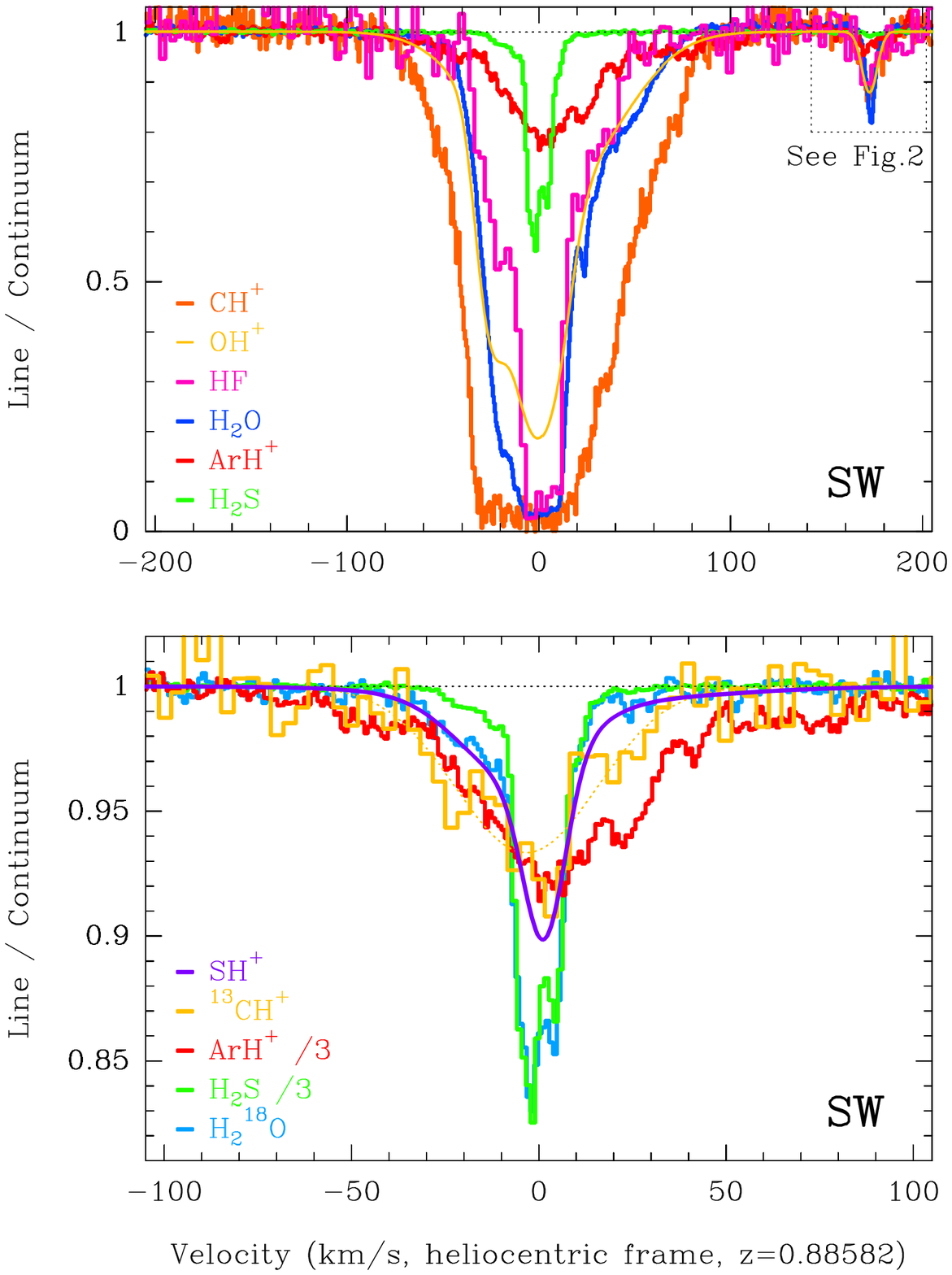} 
\caption{Absorption spectra toward the southwest image of \PKS1830: {\em top:} of the CH$^+$ $J$=1-0, OH$^+$ $N$=1-0 (deconvolved from its hyperfine structure and shown for its strongest hyperfine component), HF $J$=1-0, ortho-H$_2$O $1_{10}$-$1_{01}$, ArH$^+$ $J$=1-0, and para-H$_2$S $1_{11}$-$0_{00}$ lines; {\em bottom:} of the $^{13}$CH$^+$ $J$=1-0, SH$^+$ $N_J$=$1_2$-$0_1$ (deconvolved from its hyperfine structure and shown for its strongest hyperfine component), ArH$^+$ $J$=1-0 (opacity scaled down by a factor of three), para-H$_2$S $1_{11}$-$0_{00}$ (opacity scaled down by a factor of three), and ortho-H$_2^{18}$O $1_{10}$-$1_{01}$ lines. All spectra are normalized to the continuum level and are referenced to the heliocentric frame taking $z = 0.88582$.}
\label{fig:specSW}
\end{center} \end{figure}

Toward the SW image, the CH$^+$ spectrum shows deep and broad absorption (Fig.\,\ref{fig:specSW}), the broadest of all molecular species observed so far at millimeter wavelengths toward this source (e.g., \citealt{mul14a}). The full width at zero power exceeds 160~\kms. The line appears flat-bottomed over a velocity range of $\sim$50~\kms\ (i.e., between $-$30~\kms\ and +20~\kms), although the absorption does not exactly reach the 0 level, but implies a continuum source covering factor, $f_c \sim 97\%$. The covering factor is slightly higher than for saturated species observed before (\citealt{mul14a}). Most likely, this is due to the smaller continuum size at higher frequencies and/or to the intrinsic larger filling factor of CH$^+$, although we cannot exclude time variations (\citealt{mul08}).

The heavy saturation prevents us to derive directly the peak opacity of CH$^+$, its total column density, and chemical correlation with other species near $v=0$~\kms\ velocities. By cutting the optical depths at a threshold of $\tau =3$, we estimate a lower limit of $6.4 \times 10^{14}$~\cm-2.

\begin{figure}[h] \begin{center}
\includegraphics[width=8.8cm]{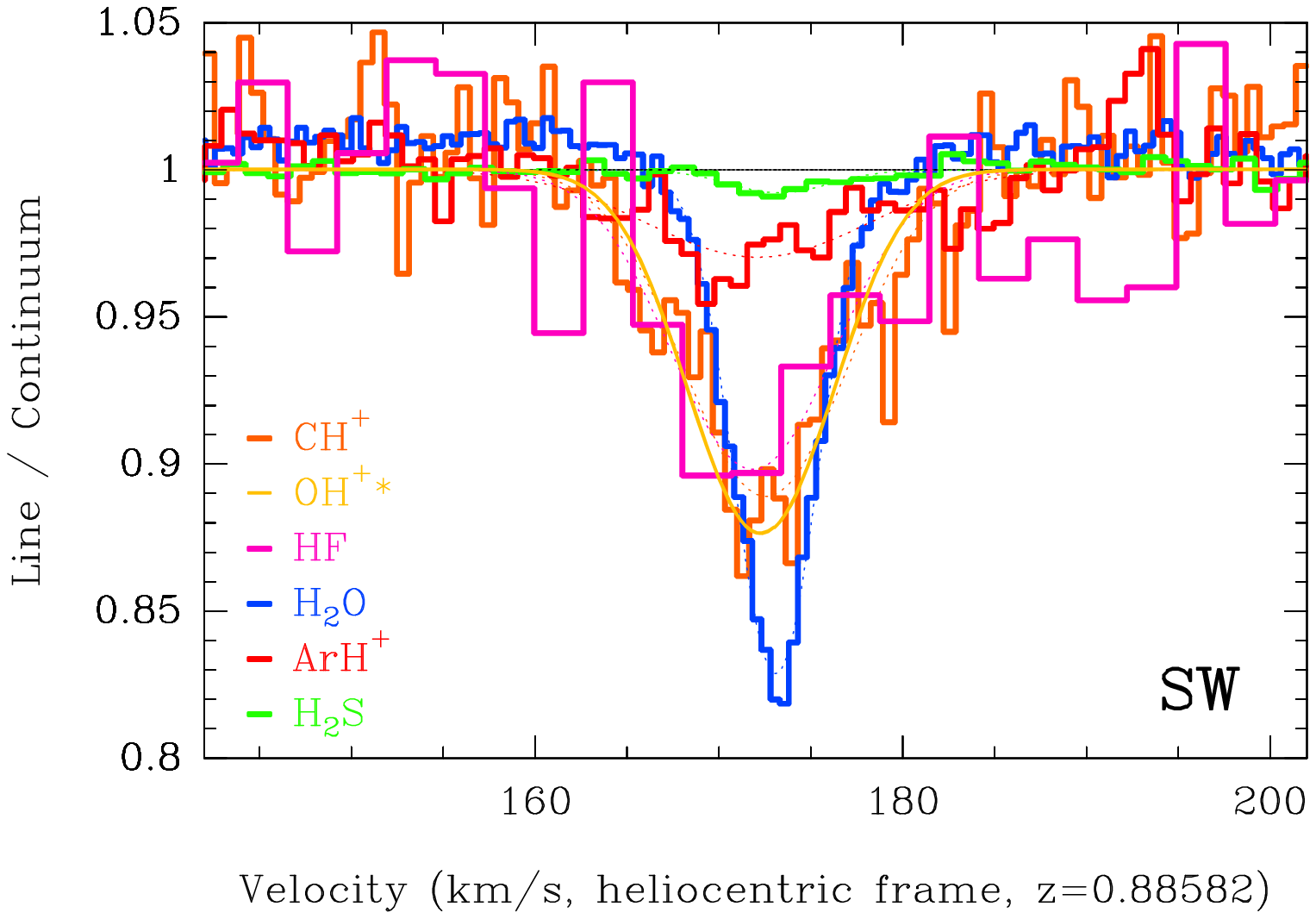}
\caption{Same as Fig.\,\ref{fig:specSW}, upper panel, zoomed on the v=+170~\kms\ velocity component toward the southwest image. The dotted curves show the best fits with one Gaussian component.}
\label{fig:spec170}
\end{center} \end{figure}

The weak $v = +170$~\kms\ component (\citealt{mul11,mul14a}) is also detected in CH$^+$, as shown in Fig.\,\ref{fig:spec170}, with an apparent peak opacity of $\sim 0.1$ and a FWHM $\sim 10$~\kms, leading to a column density $\sim 4 \times 10^{12}$~\cm-2.

The absorption from $^{13}$CH$^+$ is detected in the same tuning as CH$^+$ and barely reaches a depth of 10\% of the continuum level near $v = 0$~\kms. The hyperfine splitting of the $^{13}$CH$^+$ $J$=1-0 line is smaller than 2~MHz (rest frame), that is, spread over a velocity interval $\sim$0.7~\kms. This is much smaller than the FWHM of the line, and is thus neglected here. The total column density is about $9 \times 10^{12}$~\cm-2. 

We further discuss the CH$^+$ absorption in comparison with other species and the [CH$^+$]/[$^{13}$CH$^+$] ratio in Sec.\ref{sec:discussion}.

\subsubsection{SH$^+$ and $^{34}$SH$^+$} \label{sec:obs-SH+}

\begin{figure}[h] \begin{center}
\includegraphics[width=8.8cm]{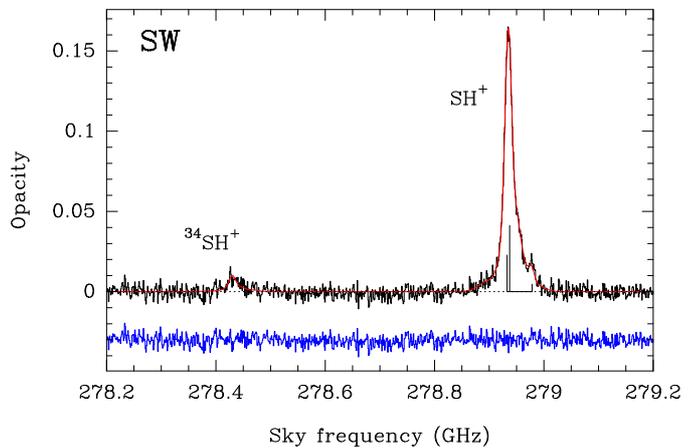} 
\caption{Opacity spectra of the SH$^+$ and $^{34}$SH$^+$ lines toward the SW image of \PKS1830, shown with their best fit model (three Gaussian components convolved with the hyperfine structure, in red) and the fit residuals (in blue, offset by -0.03). The hyperfine structure is indicated for SH$^+$. We assumed the same hyperfine structure, shifted in frequency by a constant value for $^{34}$SH$^+$ (see Sec.\,\ref{sec:obs-SH+}).}
\label{fig:fit-SH+}
\end{center} \end{figure}

The SH$^+$ absorption is detected near $v=0$~\kms\ with an optical depth $\sim 0.15$. The identification of SH$^+$ is corroborated by the presence of $^{34}$SH$^+$ at the expected frequency (see below and Sec.\,\ref{sec:spectro}). Spectra for both isotopologues, converted into opacity scale, are shown in Fig.\ref{fig:fit-SH+}. The absorption profile of SH$^+$, deconvolved from its hyperfine structure and in velocity scale, is shown in Fig.\,\ref{fig:specSW}. This profile was obtained by the fit of Gaussian velocity components convolved with the hyperfine structure. Only three Gaussian components are required to fit the profile, leaving residuals at the level of the noise, see Fig.\,\ref{fig:fit-SH+}. In particular, two Gaussian components are necessary to reproduce the large and asymmetric blue and red wings.

The companion absorption from $^{34}$SH$^+$ can also be included in the same fit. As an exercise to compare with spectroscopic calculations (see Sec.\,\ref{sec:spectro}), we assume for $^{34}$SH$^+$ the same hyperfine structure as for SH$^+$ (i.e., same splitting and relative line intensities) but shifted in frequency by a constant value $\delta$, and the same intrinsic velocity profile. We find $\delta = 952.5 \pm 1.9$~MHz, consistent with but not providing more accurate values than the frequencies listed in Table\,\ref{tab:spectro34SH+}. In this fit, we also find a ratio SH$^+$/$^{34}$SH$^+$ of $16.2 \pm 1.3$, slightly higher than the $^{32}$S/$^{34}$S ratios obtained previously from CS ($10.4 ^{+0.8} _{-0.7}$) and H$_2$S ($10.6 \pm 0.9$) isotopologues by \cite{mul06}, but significantly smaller than the Solar System value of 22 (\citealt{lod03}).

\subsection{\PKS1830\ north-east line of sight}

\subsubsection{CH$^+$ and $^{13}$CH$^+$}

\begin{figure}[h] \begin{center}
\includegraphics[width=8.8cm]{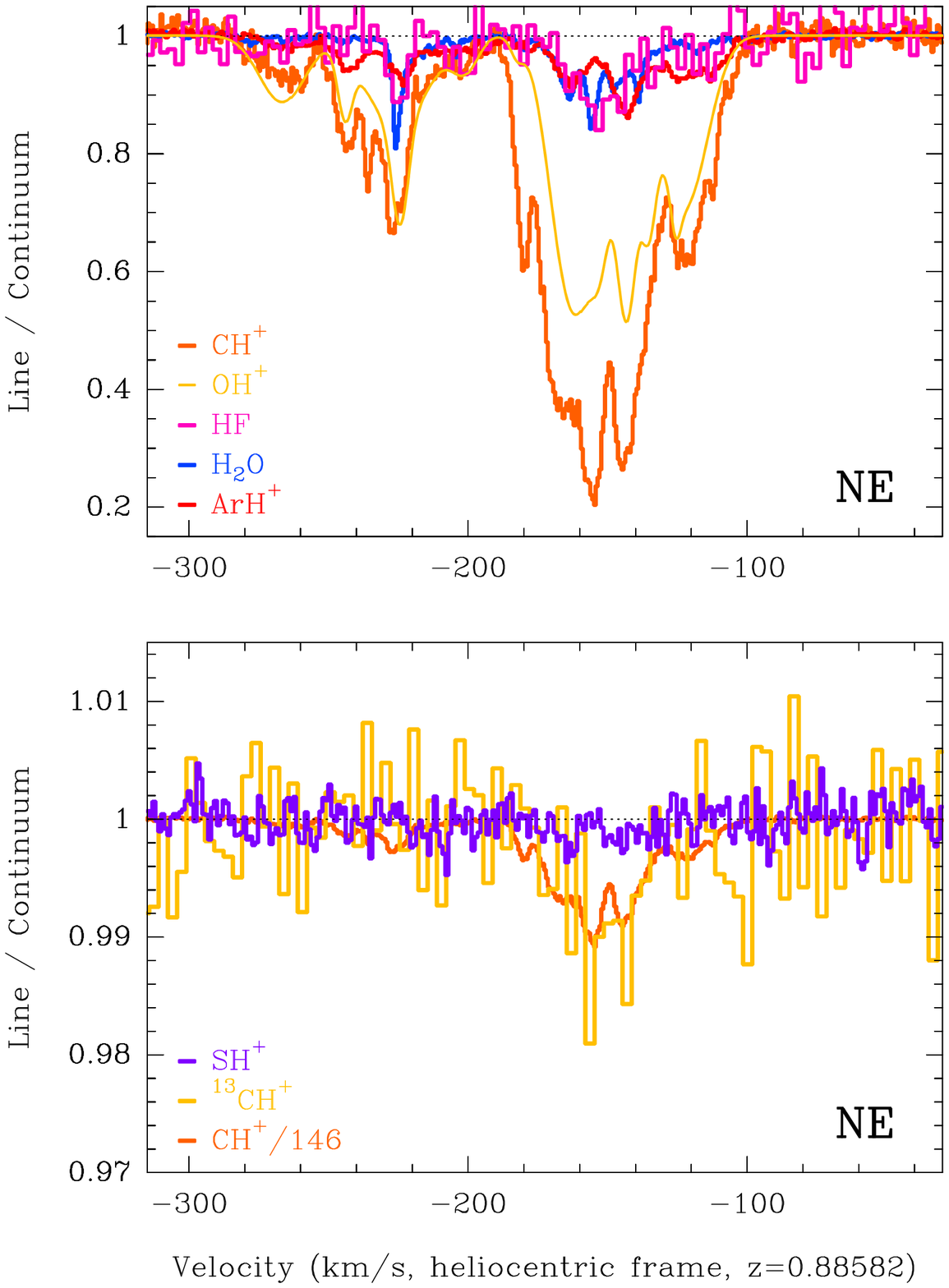} 

\caption{Absorption spectra toward the northeast image of \PKS1830: {\em top:} of the CH$^+$ $J$=1-0, OH$^+$ $N$=1-0 (deconvolved from its hyperfine structure and shown for its strongest hyperfine component), HF $J$=1-0, ortho-H$_2$O $1_{10}$-$1_{01}$, and ArH$^+$ $J$=1-0 lines; {\em bottom:} of the $^{13}$CH$^+$ $J$=1-0, SH$^+$ $N_J$=$1_2$-$0_1$ (deconvolved from its hyperfine structure and shown for its strongest hyperfine component), and CH$^+$ $J$=1-0 (opacity scaled down by a factor of 146). All spectra are normalized to the continuum level and are referenced to the heliocentric frame taking $z = 0.88582$. The $^{13}$CH$^+$ spectrum was smoothed to 3.3~\kms\ for better signal-to-noise ratio in individual channels.}
\label{fig:specNE}
\end{center} \end{figure}

\begin{figure}[h] \begin{center}
\includegraphics[width=8.8cm]{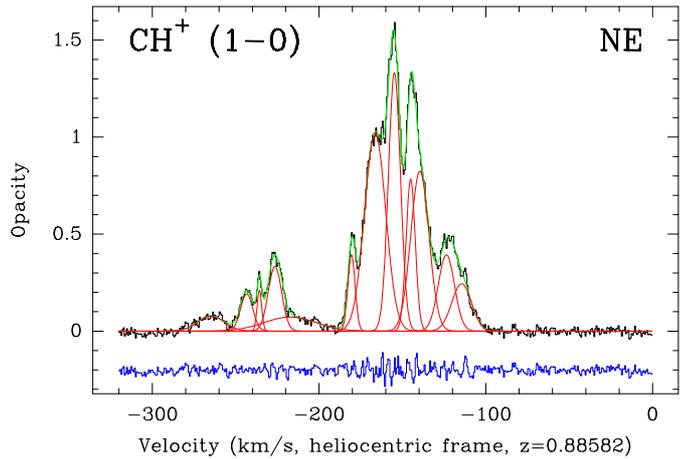} 
\caption{Fit of the CH$^+$ J=1-0 line opacity profile toward the NE image of \PKS1830. The individual Gaussian components (listed in Table\,\ref{tab:fitNE}) are marked in red and the global resulting profile is shown in green. The fit residuals are shown in blue, offset by $-$0.2 in opacity.}
\label{fig:fitNE}
\end{center} \end{figure}

The CH$^+$ NE absorption profile consists of a series of individual velocity components, with widths as small as a few \kms, spanning the large continuous interval between $-$300~\kms\ and $-$100~\kms\ (Fig.\,\ref{fig:specNE}). A fit of 12 Gaussian velocity components reproduces well the overall line opacity profile (see Fig.\,\ref{fig:fitNE} and the list of velocity components in Table~\ref{tab:fitNE}). The absorption reaches a depth of about 80\% of the continuum level at $v \sim -156$~\kms, implying that the source covering factor $f_c$ is $>$80\% for this peak absorption. Without stronger constraints, we assume $f_c = 1$ toward this image. The total CH$^+$ column density is about $1.9 \times 10^{14}$~\cm-2.

The $^{13}$CH$^+$ absorption is weakly detected. It is the first such detection for a $^{13}$C isotopologue toward the NE image of \PKS1830. It reaches a depth of $\sim$1\% of the continuum level. A simultaneous fit of the CH$^+$ and $^{13}$CH$^+$ spectra (with the same 12-Gaussian velocity components profile obtained above) yields a [CH$^+$]/[$^{13}$CH$^+$] ratio of $146 \pm 43$. The total $^{13}$CH$^+$ column density is about $1.3 \times 10^{12}$~\cm-2 along the NE line of sight.

\begin{table}[h]
\caption{Gaussian velocity components used to fit the CH$^+$ opacity profile toward the NE image of \PKS1830, as shown in Fig.\,\ref{fig:fitNE}.} \label{tab:fitNE}
\begin{center} \begin{tabular}{ccc}
\hline \hline
$v_0$ & FWHM & $\int \tau dv$ \\
(\kms) & (\kms) & (\kms) \\
\hline
$ -264.9 \pm 1.3 $ & $ 18.6 \pm  2.6 $ & $  1.6 \pm 0.3 $ \\
$ -243.4 \pm 0.3 $ & $  8.6 \pm  1.2 $ & $  1.8 \pm 0.6 $ \\
$ -235.8 \pm 0.1 $ & $  3.3 \pm  0.4 $ & $  0.7 \pm 0.1 $ \\
$ -226.5 \pm 0.2 $ & $  9.5 \pm  0.9 $ & $  3.4 \pm 0.7 $ \\
$ -217.4 \pm 9.6 $ & $ 40.0 \pm 13.3 $ & $  2.9 \pm 1.6 $ \\
$ -180.6 \pm 0.1 $ & $  5.1 \pm  0.2 $ & $  2.1 \pm 0.1 $ \\
$ -166.4 \pm 0.2 $ & $ 14.9 \pm  0.5 $ & $ 16.2 \pm 0.6 $ \\
$ -154.8 \pm 0.2 $ & $  8.5 \pm  0.4 $ & $ 12.1 \pm 1.1 $ \\
$ -145.1 \pm 0.1 $ & $  6.5 \pm  0.9 $ & $  5.4 \pm 3.1 $ \\
$ -139.6 \pm 1.8 $ & $ 12.8 \pm  3.4 $ & $ 11.2 \pm 4.4 $ \\
$ -123.6 \pm 1.4 $ & $ 11.5 \pm  2.9 $ & $  4.8 \pm 3.3 $ \\
$ -114.5 \pm 4.3 $ & $ 13.8 \pm  3.9 $ & $  3.6 \pm 2.8 $ \\
\hline
\end{tabular}
\end{center} \end{table}

\subsubsection{SH$^+$}

Despite a sensitivity better than 0.2\% of the continuum level, SH$^+$ is not detected toward the NE image. We estimate an upper limit of the SH$^+$ integrated opacity as $3 \sigma \sqrt{ \delta V \times {\rm FWHM}} = 0.036$~\kms, where we take the FWHM=42~\kms\ determined from the fit of the CH$^+$ absorption with only one Gaussian velocity component, and where $\sigma$ is the standard deviation of the spectrum and $\delta V$ the velocity resolution. This corresponds to an upper limit of $3.2 \times 10^{11}$~\cm-2 for the column density.

The abundance ratio between the SW and NE lines of sight, provided excitation conditions are comparable, is then larger than 120, which is much higher than for other hydrides like ArH$^+$, OH$^+$, H$_2$O$^+$, H$_2$Cl$^+$, and CH$^+$ (see Table\,\ref{tab:ncol} and Sec.\,\ref{sec:gamma}).

\section{Discussion} \label{sec:discussion}

\subsection{Comparison with other species}

\subsubsection{Time variations}

As previously shown (e.g., \citealt{mul08,sch15}), we have to account for the effect of time variations for the comparison of spectra taken at different epochs toward \PKS1830, with an observed timescale of the order of a month to years. The lensing geometry of the system is such that morphological changes in the background quasar, for example due to the emission of new plasmons in the (possibly helical) jet (\citealt{gar97,jin03,nai05}), can cause a slightly varying illumination of clouds in the absorber and thereby changes in the absorption line profiles toward both lensed images. 

We do not expect micro-lensing, e.g., by a stellar-mass object in the lens plane, to cause effective variability, because the continuum emitting size of the quasar at mm wavelengths is still several orders of magnitude larger than the corresponding Einstein radius, $\theta_{\rm E, micro} \sim 2 \mu$as. A transverse velocity $\sim 1000$~\kms\ in the lens plane would convert into an apparent drift of $\sim 0.5$~mpc within one year. The timescale associated with milli-lensing by an object of $>>$ 1~M$_\odot$ would thus be of the order of weeks or longer.

ArH$^+$ (\citealt{hmul15a}) and HF (\citealt{kaw16}) were observed by ALMA the day before CH$^+$ in May 2015, while OH$^+$ was observed about two weeks later (\citealt{mul16}). In May and July 2014, the ground-state transitions of ortho-H$_2^{18}$O were observed in the same tuning as SH$^+$, and another sulfur-bearing molecule, para-H$_2$S, was observed within one month. Within such short time intervals, we do not expect significant variations of the absorption profiles.

\cite{mul16} discuss the temporal variations of H$_2$O, CH, and H$_2$O$^+$ profiles between the different ALMA observing sessions in 2014, 2015, and 2016. The major changes occur in the blue and red wings of the saturated line of H$_2$O toward the SW image, with an increase of $\sim$50\% of the integrated opacities in the velocity ranges $-60$ to $-20$~\kms\ and +20 to +80~\kms\ between 2014 and 2016. For CH, the total integrated opacity varies by less than 6\%. Toward the NE image, the total integrated opacity of the water line varies by less than 15\% in this period. Different fine structure transitions of H$_2$O$^+$, observed between 2014 and 2015, do not show significant differences.

These checks allow us to link the profiles of the different species observed with ALMA between 2014 and 2015, and suggest little evolution of the absorption line profiles toward the two images of \PKS1830\ during this period.

\cite{all17} recently published a study of the long-term variability of the 21~cm H\,I line toward \PKS1830. They compare the H\,I profiles observed in 1996 by \cite{che99} and 1999 by \cite{koo05} with new data obtained in 2014-2015. In stark contrast with the molecular variability at millimeter wavelengths, they find only marginal variations of at most few percent within this long time span. This is consistent with the fact that the continuum at cm wavelengths is much more extended, smoothing away all variations. In the absence of better estimates of H\,I column densities, we use those derived from the source kinematical model by \cite{koo05} and given in Table\,\ref{tab:ncol} (see also \citealt{mul16}).

\subsubsection{Column density ratios}

\begin{figure}[h] \begin{center}
\includegraphics[width=8.8cm]{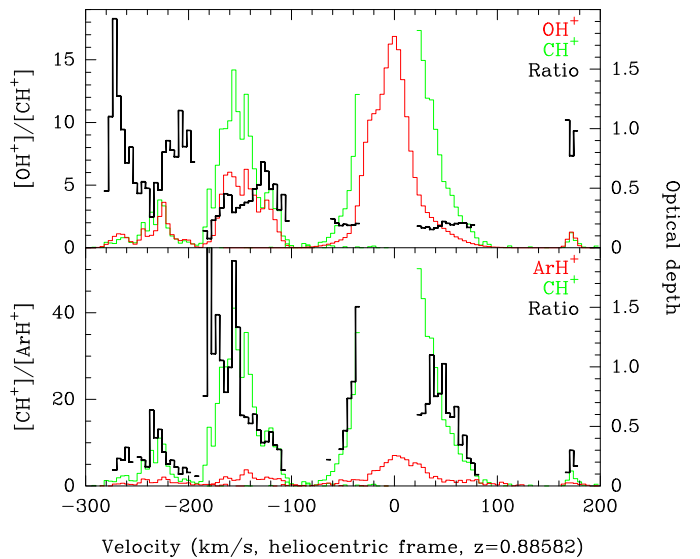} 
\caption{ {Column density ratios of [OH$^+$]/[CH$^+$] ({\em top}) and [CH$^+$]/[ArH$^+$] ({\em bottom}) toward the NE ($v < -100$~\kms) and SW ($v > -100$~\kms) lines of sight. The ratios are shown in black curves. The green and red curves are optical depths and translate into column densities with factors $2.85 \times 10^{12}$, $20.0 \times 10^{12}$,  and $1.70 \times 10^{12}$ \cm-2\,km$^{-1}$\,s for CH$^+$, OH$^+$, and ArH$^+$, respectively. Channels with SNR~$< 3$ or opacity larger than 2 were flagged prior to calculate the ratios.}
}
\label{fig:ratio-CH+}
\end{center} \end{figure}

Among the species used for comparison in this paper, only OH$^+$, and ArH$^+$ have wings sufficiently broad toward the SW image to avoid the saturated region in the CH$^+$ spectrum and are strong enough toward the NE image, for determining column density ratios with CH$^+$. These ratios are shown in Fig.\,\ref{fig:ratio-CH+} for both lines of sight. The [OH$^+$]/[CH$^+$] ratio is rather constant, $=1.8 \pm 0.2$, in the wings of the SW absorption ($v = -60$ to $-$40~\kms\ and +20 to +80 \kms), but varies significantly between $\sim$1 and 18 for all other velocities of the NE absorption ($v = -300$ to $-$100~\kms). It is $\sim 9$ for the $v=+170$~\kms\ component toward the SW image. On the other hand, the [CH$^+$]/[ArH$^+$] ratio is varying abruptly between $\sim 2$ and 60, and never stablizes to a constant value. We note a clear trend that when [OH$^+$]/[CH$^+$] is low, [CH$^+$]/[ArH$^+$] is high, and vice versa, one possible explanation being that CH$^+$ traces gas with higher \fH2\ than ArH$^+$ and OH$^+$.

Assuming that the [OH$^+$]/[CH$^+$] ratio is constant over the SW absorption, i.e., also in the saturated region of the CH$^+$ absorption, we can perform a simultaneous fit of the CH$^+$, $^{13}$CH$^+$, and OH$^+$ spectra, using Eq.\,\ref{eq:abs_function} with a single intrinsic velocity profile for all species. The free parameters of the fit are thus the centroid velocity, full width at half maximum (FWHM), and integrated opacity for each Gaussian velocity component, the source covering factor, $f_c$ (we assume the same for each species), and the scaling factors between species. The best fit with three Gaussian components yields $fc= 97.8 \pm 0.3$~\%, [CH$^+$]/[$^{13}$CH$^+$]~$= 97 \pm 6$, and [OH$^+$]/[CH$^+$]~$=1.7 \pm 0.1$. It reproduces well the spectra and leaves residuals close to the noise level. For simplicity, we assume Gaussian statistics on the errors in the fit. From this, we derive a new ``saturation-corrected'' total column density for CH$^+$ of $9.7 \times 10^{14}$~\cm-2. Using the H\,I and H$_2$ column densities from Table\,\ref{tab:ncol}, we finally obtain the CH$^+$ abundance relative to total hydrogen: N(CH$^+$)/N(H) = N(CH$^+$)/(N(H\,I)+2$\times$N(H$_2$))~$=2 \times 10^{-8}$ and $4 \times 10^{-8}$ along the SW and NE lines of sight, respectively, comparable to typical values found in the Milky Way (e.g., \citealt{fal10,men11, god12}).

\begin{figure}[h] \begin{center}
\includegraphics[width=8.8cm]{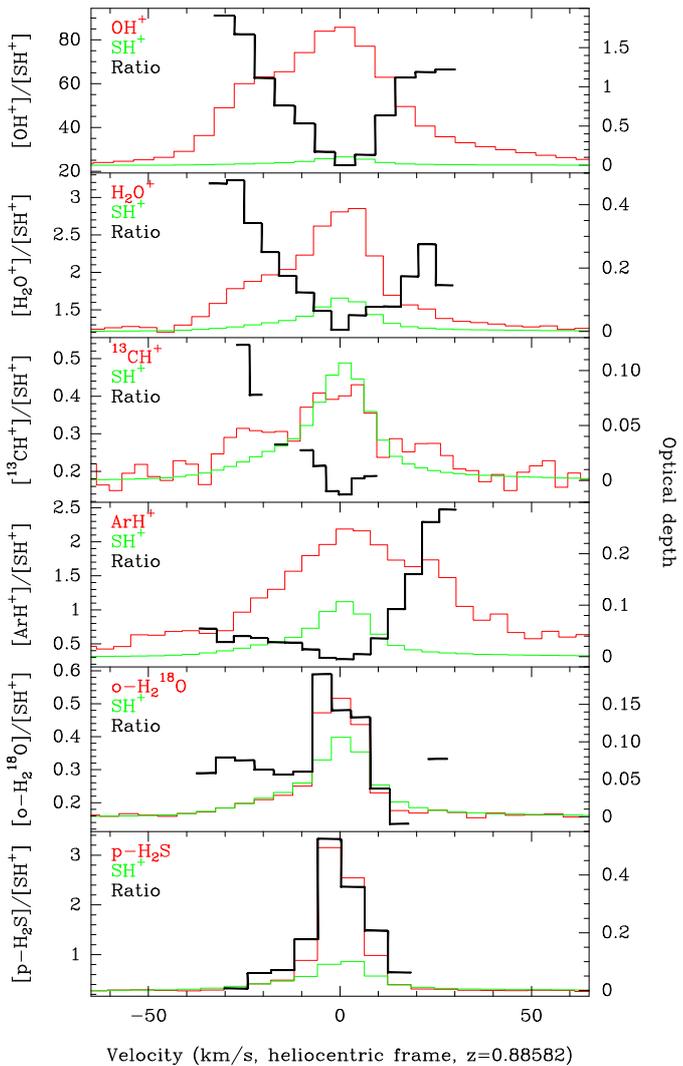} 
\caption{Column density ratios ({\em in black}) of several species ({\em in red}) relative to SH$^+$ ({\em in green}) toward the SW image of \PKS1830. Opacity spectra were smoothed by five channels to improve the signal-to-noise ratio, SNR, in individual channels. The column density ratios are calculated only where both species have opacity SNR larger than five.}
\label{fig:ratio-SH+}
\end{center} \end{figure}

In Fig.\,\ref{fig:ratio-SH+}, we show the column density ratios of SH$^+$ relative to other species discussed in this work, for the SW line of sight. These ratios show a clear pattern with different species and velocity ranges. At $|v| < 10$~\kms, the ratios [X]/[SH$^+$] go to a minimum for X=OH$^+$, H$_2$O$^+$, $^{13}$CH$^+$, and ArH$^+$, all hydrides preferentially tracing gas with low molecular fraction. The ratios increase by a factor two or more in the wings, for $|v| > 10$~\kms. For X=ortho-H$_2^{18}$O and para-H$_2$S, the ratios show a reverse trend in the velocity interval $|v| < 10$~\kms, increasing toward the line center and dropping by a factor two or more in the wings. The $|v| < 10$~\kms\ velocity interval is dominated by translucent gas with a relatively high density of a few $10^3$~cm$^{-3}$, as determined from the excitation of NH$_3$ (\citealt{hen08}) and several other species (\citealt{hen09, mul13}). In the line wings, where the low-\fH2\ tracers show enhanced absorption, the density is likely lower and the gas more diffuse. Hence, we can conclude that SH$^+$ behaves intermediate between tracers of low- and high-molecular fraction, in agreement with the analysis by \cite{neu15} along Galactic sightlines.

Taking the total column densities, we find a large difference in the [CH$^+$]/[SH$^+$] ratios between the two lines of sight toward \PKS1830: $\sim 25$ toward the SW image and $> 600$ toward the NE. This agrees with the correlation derived by \cite{god12} in the Milky Way, that shows a higher N(CH$^+$)/N(SH$^+$) ratio in regions with high N(CH$^+$)/N(H). However, the difference of N(CH$^+$)/N(H) between the two \PKS1830\ sightlines is only a factor of two, when the [CH$^+$]/[SH$^+$] varies by a factor $> 24$. The [CH$^+$]/[SH$^+$] ratios collected by \cite{god12} show a large scatter of more than two orders of magnitude. They interpret these wild variations as linked to the ion-drift velocity in the turbulent dissipation regions model, therefore the amount of suprathermal energy injected in the system, and to the difference in formation endothermicities between the two species.

\subsection{Comparison of the two lines of sight toward \PKS1830} \label{sec:gamma}

CH$^+$, SH$^+$, and their $^{13}$C- and $^{34}$S- isotopologues are detected toward the SW image of \PKS1830. This line of sight is known to contain nearly 50 molecular species, not counting isotopologues, with an H$_2$ column density of about $2 \times 10^{22}$~\cm-2\ (see, e.g., \citealt{mul11,mul14a}). At $v \sim 0$~\kms, it mostly consists of translucent clouds at moderate density of the order of $10^3$~cm$^{-3}$ and with a kinetic temperature of $\sim$80~K (\citealt{hen09, mul13}). In contrast, the NE line of sight consists of more diffuse gas (see e.g., \citealt{mul08,mul14b,mul16}), although the physical conditions are not as well constrained as toward the SW image.

Previously, four species have been found to have enhanced abundances in the NE line of sight, with respect to the SW line of sight: H$_2$Cl$^+$ (\citealt{mul14b}), ArH$^+$ (\citealt{hmul15a}), and OH$^+$ and H$_2$O$^+$ (\citealt{mul16}). These species are also known to trace gas with low molecular fraction (\citealt{ger16}). \cite{mul16} noticed that the total column density ratio between the SW and NE images (hereafter $\gamma_{\rm SW/NE}$) for a given species is a good indicator of what kind of gas it preferentially traces. We list in Table\,\ref{tab:ncol} the column densities of several species observed toward both lines of sight, as well as their $\gamma_{\rm SW/NE}$ ratios. When ordered with increasing value of $\gamma_{\rm SW/NE}$, there is a clear trend with tracers of increasing \fH2.

In this picture, CH$^+$, with a $\gamma_{\rm SW/NE} = 5$, is intermediate between the low \fH2 tracers, e.g., OH$^+$ and H$_2$O$^+$ (tracing \fH2 of a few percent) and high \fH2 tracers, such as CH and HF, for which $\gamma_{\rm SW/NE} \gtrsim 20$. On the other hand, SH$^+$, with its non detection in the NE line of sight, yields a lower limit $\gamma_{\rm SW/NE} > 120$, a rather extreme ratio among the species listed, either suggesting that SH$^+$ is found in gas with relatively high-\fH2 or that the physical conditions in the NE line of sight are not favorable to the formation of SH$^+$. By comparison, \cite{god12} find molecular fractions with a large scatter $0.04 < \fH2 < 1$ for the diffuse gas seen in CH$^+$ and SH$^+$ absorption along Galactic sightlines, with an average value of 10\%.

In their recent chemical predictions, \cite{neu16} find a sequence ArH$^+$--HCl$^+$--H$_2$Cl$^+$--OH$^+$--H$_2$O$^+$ where the five ions reach their peak abundance at increasing molecular fraction (or visual extinction). ArH$^+$ is confirmed to be a unique tracer of almost purely atomic gas (i.e., even better than H\,I), peaking at molecular fraction $10^{-5}-10^{-2}$ (see also \citealt{sch14}). OH$^+$ and H$_2$O$^+$ are found to reside primarily in gas with $\fH2 \sim 0.01-0.1$, with the chlorine-bearing ions tracing intermediate $\fH2 \lesssim 0.1$. Although not perfect, the agreement between the $\gamma_{\rm SW/NE}$-classification from the two \PKS1830\ sightlines and the chemical predictions is remarkable.

Finally, we note that the [CH$^+$]/[CH] ratio in the NE line of sight is higher ($\sim 5$) than in the SW ($\sim 1$). In the Milky Way, an elevated [CH$^+$]/[CH] ratio suggests a larger contribution of the CH$^+$ chemistry to the formation of CH, in contrast to regions where relatively more CH arises from quiescent chemistry, as discussed by \cite{fed97} and \cite{por14}.

\begin{table*}[ht]
\caption{Total column densities of various species along the SW and NE lines of sight toward \PKS1830. Species are ordered with increasing SW/NE column density ratios, $\gamma_{\rm SW/NE}$. Note that time variations might affect the comparison of data observed far apart in time.} \label{tab:ncol}
\begin{center} \begin{tabular}{cccccc}

\hline \hline
Species & Year of       & \multicolumn{2}{c}{Column densities (\cm-2)} & $\gamma_{\rm SW/NE}$ & \fH2 \\
        & observation & SW & NE &  & \\
\hline
H\,I       & 1996-2015 $^{(a)}$ & $1.3 \times 10^{21}$ & $2.5 \times 10^{21}$ & 0.5 &  $\sim$1\% -- $>$50\% \\
ArH$^+$     & 2015 $^{(b)}$ & $2.7 \times 10^{13}$ & $1.3 \times 10^{13}$ & 2.1 &  $10^{-5} - 10^{-2}$ $^{(j)}$ \\
OH$^+$      & 2015 $^{(c)}$ & $1.6 \times 10^{15}$ & $7.6 \times 10^{14}$ & 2.2 &  a few \% \\
H$_2$Cl$^+$ & 2012 $^{(d)}$ & $1.4 \times 10^{13}$ & $3.7 \times 10^{12}$ & 3.8 &  \\
H$_2$O$^+$  & 2014 $^{(c)}$ & $2.7 \times 10^{14}$ & $7.0 \times 10^{13}$ & 3.9 &  a few \% \\
CH$^+$      & 2015 $^{(e)}$ & $9.7 \times 10^{14}$ & $1.9 \times 10^{14}$ & 5.1 &  \\
$^{13}$CH$^+$ & 2015 $^{(e)}$ & $9.8 \times 10^{12}$   & $1.3 \times 10^{12}$ & 7.5 & \\
HF         & 2015 $^{(f)}$ & $>3.4 \times 10^{14}$ & $1.8 \times 10^{13}$ & $> 19$ &  \\
CH         & 2012 $^{(g)}$ & $7.7 \times 10^{14}$ & $3.5 \times 10^{13}$ & 22. &  $\sim$10\% -- 100\% \\
HCO$^+$    & 2009 $^{(h)}$ & $1.8 \times 10^{14}$ & $7.5 \times 10^{12}$ & 23. &  \\
HCN        & 2009 $^{(h)}$ & $3.0 \times 10^{14}$ & $7.3 \times 10^{12}$ & 42. &  \\
C$_2$H     & 2009 $^{(h)}$ & $1.2 \times 10^{15}$ & $2.9 \times 10^{13}$ & 43. &  \\
c-C$_3$H$_2$ & 2009 $^{(h)}$ & $5.3 \times 10^{13}$ & $1.2 \times 10^{12}$ & 44. &  \\
N$_2$H$^+$ & 2009 $^{(h)}$ & $2.3 \times 10^{13}$ & $< 0.3 \times 10^{12}$ & $>70$ &  \\
o-H$_2^{18}$O & 2014 $^{(e)}$ & $1.5 \times 10^{13}$ & $< 1.8 \times 10^{11}$ & $> 82$ &  \\
HNC        & 2009 $^{(h)}$ & $1.0 \times 10^{14}$ & $1.2 \times 10^{12}$ & 87 &  \\
SH$^+$ & 2014 $^{(e)}$ & $3.9 \times 10^{13}$ & $< 3.2\times 10^{11}$ & $> 120$ &  \\
p-H$_2$S  & 2014 $^{(e)}$ & $6.6 \times 10^{13}$ & $< 4.8\times 10^{11}$ & $> 140$ &  \\
\hline
H$_2$      & 2012 $^{(g,i)}$ &  $\sim 2 \times 10^{22}$ & $\sim 1 \times 10^{21}$ & $\sim 20$ &  \\
\hline
\end{tabular}
\tablefoot{ $(a)$ \cite{che99, koo05,all17}; $(b)$ \cite{hmul15a}; $(c)$ \cite{mul16}; $(d)$ \cite{mul14b}; $(e)$ this work; $(f)$ \cite{kaw16}; $(g)$ \cite{mul14a}; $(h)$ \cite{mul11}; $(i)$ The column densities of H$_2$ were derived using CH and H$_2$O as proxies (\citealt{mul14a}); $(j)$ \cite{sch14,neu16}.}
\end{center} \end{table*}

\subsection{$^{12}$C/$^{13}$C ratio} \label{sec:12/13C}

Because of its relatively large value ($\sim 60$ in the local ISM, \citealt{luc98}), the interstellar $^{12}$C/$^{13}$C ratio can be difficult to measure, either due to saturation of the main $^{12}$C-species or sensitivity issues with the detection of the $^{13}$C-isotopologues. In addition, the interpretation of the [$^{12}$CX]/[$^{13}$CX] abundance ratio can be complicated by fractionation and/or selective photodissociation. From optical-line absorption studies in the Milky Way, \cite{rit11} find that the [$^{12}$CH$^+$]/[$^{13}$CH$^+$] ratio does not deviate from the $^{12}$C/$^{13}$C isotopic ratio, as expected if the molecule forms via energetic processes.

Our measurement of [$^{12}$CH$^+$]/[$^{13}$CH$^+$] toward the SW image is indirect, using the [OH$^+$]/[CH$^+$] ratio determined from the line wings to link CH$^+$ in the saturated velocity interval near $v = 0$~\kms\ to $^{13}$CH$^+$. We derive [$^{12}$CH$^+$]/[$^{13}$CH$^+$]~$= 97 \pm 6$ in a combined fit, which would imply a large optical depth $\sim 5-10$ for the peak of the CH$^+$ absorption. This [$^{12}$CX]/[$^{13}$CX] ratio is much higher than the previous measurements $\sim 30-40$ using HCO$^+$, HCN, and HNC (\citealt{mul06,mul11}). Non detection of the $^{13}$C-variants for H$_2$CO and C$_2$H sets limits to the ratio significantly higher than 40 (\citealt{mul11}), and may suggest fractionation issues (see e.g., \citealt{rou15}). Alternatively, it could be possible that the isotopic ratio is different due to incomplete mixing between different gas components: low- {\em vs} high-\fH2 gas or molecular gas that has been more enriched by recent stellar formation {\em vs} a more highly disturbed component of the ISM.

For the first time toward the NE image, we are able to measure a [$^{12}$CX]/[$^{13}$CX] ratio. From a combined fit of the CH$^+$ and $^{13}$CH$^+$ spectra, we obtain [$^{12}$CH$^+$]/[$^{13}$CH$^+$]~$= 146 \pm 43$. In this fit, we set the ratio as a free parameter and assume Gaussian statistics on the uncertainties. As for the $^{36}$Ar/$^{38}$Ar ratio (\citealt{hmul15a}), we find a slight difference between the two lines of sight, suggesting that the NE lines of sight, intercepting the absorber at a galactocentric radius of $\sim 4$~kpc compared to $\sim 2$~kpc for the SW line of sight, might be composed of less processed material.

\section{Summary and conclusions} \label{sec:conclusions}

We report ALMA observations of CH$^+$, SH$^+$, and their $^{13}$C- and $^{34}$S-isotopologues along two independent lines of sight, with different physico-chemical properties, across the $z$=0.89 absorber toward \PKS1830.

CH$^+$ shows deep absorption spanning a velocity range of $\sim 200$~\kms\ along both sightlines, with $^{13}$CH$^+$ also detected, albeit weakly, along both. In constrast, SH$^+$ is only detected toward the SW line of sight, characterized by an higher average molecular fraction \fH2. We report the first interstellar detection of $^{34}$SH$^+$ in the same line of sight.

The [CH$^+$]/[SH$^+$] column density ratios differ widely between the two sightlines, $\sim 25$ in the SW, and $>600$ in the NE. This suggests, in agreement with previous observations, that SH$^+$ resides in gas with high \fH2 ($\gtrsim 10$\%). Alternatively, the difference of column density ratios might be due to the difference of formation endothermicity between the two species, with physical conditions not as favorable to the formation of SH$^+$ in the NE line of sight.

We suggest that the total column density ratios between the two lines of sight for a given species is a good indicator of the molecular fraction where the species primarily resides. In this picture, CH$^+$ is placed in gas with $\fH2 \gtrsim 10$\%, i.e., with similar or possibly slightly higher \fH2 than for OH$^+$ and H$_2$O$^+$. The CH$^+$ $J$=1-0 line is the most heavily saturated line among all species previously observed in this source, and as such, appears as the best tracer of diffuse gas in terms of detectability, in particular at high redshift.

The detection of $^{13}$CH$^+$ allows us to estimate [$^{12}$CH$^+$]/[$^{13}$CH$^+$] $\sim 100$ and $\sim 150$, toward the SW and NE sightlines, respectively. The SW value is larger than any previous [$^{12}$CX]/[$^{13}$CX] ratios determined from HCO$^+$, HCN, and HNC. The even larger (although somehow uncertain) value derived in the NE line of sight, intercepting the disk of the absorber at a larger galactocentric radius, suggests that the material might be less processed by stellar nucleosynthesis.

This work shows that with the unprecedented sensitivity and frequency coverage of ALMA, we now have the opportunity to follow up the recent breakthroughs in our understandings of the physics and chemistry in the ISM of our Milky Way, with observations of hydrides in distant galaxies, and of the $z$=0.89 absorber toward \PKS1830, in particular.

\begin{acknowledgement}
We thank the referee for useful comments and corrections. This paper makes use of the following ALMA data: ADS/JAO.ALMA\#2012.1.00056.S and \#2013.1.00020.S. ALMA is a partnership of ESO (representing its member states), NSF (USA) and NINS (Japan), together with NRC (Canada) and NSC and ASIAA (Taiwan) and KASI (Republic of Korea), in cooperation with the Republic of Chile. The Joint ALMA Observatory is operated by ESO, AUI/NRAO and NAOJ. This research has made use of NASA's Astrophysics Data System.
\end{acknowledgement}


\begin{appendix}

\section{Complementary Laboratory Data} \label{app:SH+}

We have evaluated the SH$^+$ spectroscopic parameters in the present work. Transition frequencies with microwave accuracy exist only for the $N$=1-0 rotational transition; all the following data, including multiple determinations, were used in the present fit. \citet{sav04} reported data for the $J$=0-1 and $J$=2-1 fine structure (FS) components. \citet{hmul14} analyzed ALMA data of the Orion Bar region and showed that the $J$=0-1 datum of \citet{sav04} was in error by several megahertz and reported both hyperfine structure (HFS) components. \citet{hal15} obtained improved transition frequencies for all three FS components and confirmed the findings from ALMA observations \citep{hmul14}. Even though \citet{hal15} only fit their own data, their fit displayed large residuals (up to 204~kHz) between measured transition frequencies and those calculated from their spectroscopic parameters much larger than the experimental 50~kHz. Exchanging the transition frequency of the $J$=1-1, $F$=0.5-0.5 HFS component at 683359.227~MHz with that of a nearby line at 683360.577~MHz yielded residuals of less than 20~kHz on average. Consequently, we employed the latter frequency in our fits. \citet{bro09} derived extrapolated zero-field frequencies from laser magnetic resonance data \citep{hov87}; in addition, rovibrational data \citep{bro86,civ89} were also used in the fit. The resulting parameters are given in Table~\ref{tab:spec-parameter}. We summarize for convenience the $N$=1-0 transition frequencies determined with microwave accuracy in Table~\ref{tab:N_1-0}.

\begin{table}
\begin{center}
\caption{Spectroscopic parameters $^{a}$ (MHz, cm$^{-1}$) of sulfanylium, SH$^+$.}
\label{tab:spec-parameter}
\begin{tabular}[t]{lr@{}l}
\hline \hline
Parameter                     & \multicolumn{2}{c}{Value} \\
\hline
$Y_{10}$ $^b$                  &    2\,547&.4950~(104) \\
$Y_{20}$ $^b$                  &     $-$49&.4296~(90)  \\
$Y_{30}$ $^b$                  &         0&.2098~(30)  \\
$Y_{40} \times 10^3$ $^b$      &     $-$16&.03~(34)    \\
$Y_{01}$                      &  278\,095&.11~(36)    \\
$Y_{11}$                      & $-$8\,577&.42~(85)    \\
$Y_{21}$                      &        16&.18~(32)    \\
$Y_{02}$                      &     $-$14&.7431~(68)  \\
$Y_{12} \times 10^3$          &       123&.7~(26)     \\
$Y_{03} \times 10^3$          &         0&.46 $^c$     \\
$\lambda _{00}$               &  171\,489&.6~(57)     \\
$\lambda _{10}$               &    $-$473&.0~(145)    \\
$\lambda _{20}$               &     $-$78&.6~(67)     \\
$\lambda _{01}$               &      $-$1&.35~(17)    \\
$\gamma _{00}$                & $-$5\,036&.46~(89)    \\
$\gamma _{10}$                &       116&.5~(20)     \\
$\gamma _{20}$                &         3&.52~(64)    \\
$\gamma _{01}$                &         0&.433~(35)   \\
$b_{F,0}(^1$H)                &     $-$56&.840~(32)   \\
$b_{F,1}(^1$H$)-b_{F,0}(^1$H) &      $-$3&.51~(79)    \\
$c(^1$H)                      &        33&.482~(134)  \\

\hline
\end{tabular}\\[2pt]
\end{center}
\tablefoot{$(a)$ Numbers in parentheses are one standard deviation in units of the least significant digits. 
$(b)$ In units of cm$^{-1}$. $(c)$ Kept fixed to value derived by \citet{bro09}.}

\end{table}


\begin{table}
\caption{Quantum numbers $J$ and $F$, frequencies (MHz), uncertainties unc. (kHz), and residuals o$-$c (kHz) between observed rest frequencies obtained with microwave accuracy used in the present fit and those calculated from the present set of spectroscopic parameters of the $N$=1-0 transition  of sulfanylium, SH$^+$, and notes on the source.} 
\label{tab:N_1-0}
\begin{center}
\begin{tabular}{ccr@{}lrrc}
\hline\hline
$J' - J''$ & $F' - F''$ & \multicolumn{2}{c}{Frequency} & \multicolumn{1}{c}{unc.} 
 & \multicolumn{1}{c}{o$-$c} & Note \\[1pt]
\hline

$0-1$ & $0.5-0.5$ & 345858&.271 &  50 &    24 & $a$ \\
$0-1$ & $0.5-0.5$ & 345858&.270 & 200 &    23 & $b$ \\
$0-1$ & $0.5-1.5$ & 345944&.420 &  50 & $-$21 & $a$ \\
$0-1$ & $0.5-1.5$ & 345944&.350 & 200 & $-$50 & $b$ \\
$2-1$ & $1.5-0.5$ & 526038&.793 &  50 &     7 & $a$ \\
$2-1$ & $1.5-0.5$ & 526038&.722 &  75 & $-$64 & $c$ \\
$2-1$ & $2.5-1.5$ & 526048&.023 &  50 &    23 & $a$ \\
$2-1$ & $2.5-1.5$ & 526047&.947 &  75 & $-$53 & $c$ \\
$2-1$ & $1.5-1.5$ & 526124&.951 &  50 &    12 & $a$ \\
$2-1$ & $1.5-1.5$ & 526124&.976 &  75 &    37 & $c$ \\
$1-1$ & $1.5-0.5$ &(683334&.681)&  67 &   $-$ & $d$ \\
$1-1$ & $0.5-0.5$ & 683360&.577 &  50 &     5 & $e$ \\
$1-1$ & $1.5-1.5$ & 683420&.835 &  50 &     2 & $a$ \\
$1-1$ & $0.5-1.5$ &(683446&.724)&  67 &   $-$ & $d$ \\

\hline
\end{tabular}\\[2pt]
\end{center}
\tablefoot{$(a)$ \citet{hal15}. $(b)$ \citet{hmul14}. $(c)$ \citet{sav04}. 
  $(d)$ Frequencies and uncertainties calculated from the spectroscopic parameters 
  in Table~\ref{tab:spec-parameter}. $(e)$ This work and \citet{hal15}.}

\end{table}

The reevaluated SH$^+$ spectroscopic parameters will be used to create an updated 
entry for the CDMS catalog \citep{CDMS01,CDMS05}.

\end{appendix}

\end{document}